\documentstyle[preprint,revtex]{aps}
\tightenlines
\mediumtext
\begin{document}
\draft
\def\slepton{\widetilde \ell}
\def\sl{{\widetilde \ell}^{-}}
\def\slb{{\widetilde \ell}^{+}}
\def\slr{\slepton_{R}}
\def\sll{\slepton_{L}}
\def\squark{\widetilde q}
\def\msl{m_{\slepton}}
\def\msq{m_{\squark}}
\def\msll{m_{\sll}}
\def\mslr{m_{\slr}}
\def\photino{\widetilde \gamma}
\def\zino{{\widetilde{Z}}}
\def\wino{{\widetilde{W}}}
\def\sfermion{\widetilde f}
\def\sf{\widetilde f}
\def\gluino{\widetilde g}
\def\sfb{\widetilde{\overline f}}
\def\sneutrino{\widetilde \nu}
\def\sele{\widetilde e}
\def\ser{\sele_{R}}
\def\sel{\sele_{L}}
\def\sql{\squark_{L}}
\def\sqr{\squark_{R}}
\def\msf{m_{\sfermion}}
\def\msn{m_{\sneutrino}}
\def\msg{m_{\gluino}}
\def\smu{\widetilde \mu}
\def\stau{\widetilde \tau}
\def\staul{\stau_{L}}
\def\staur{\stau_{R}}
\def\su{\widetilde{u}}
\def\sd{\widetilde{d}}
\def\sc{\widetilde{c}}
\def\ss{\widetilde{s}}
\def\sul{\su_{L}}
\def\sdl{\sd_{L}}
\def\sur{\su_{R}}
\def\sdr{\sd_{R}}
\def\sclr{\sc_{L,R}}
\def\sslr{\ss_{L,R}}
\def\sulr{\su_{L,R}}
\def\sdlr{\sd_{L,R}}
\def\msul{m_{\sul}}
\def\msur{m_{\sur}}
\def\msdl{m_{\sdl}}
\def\msdr{m_{\sdr}}
\def\msclr{m_{\sclr}}
\def\msulr{m_{\sulr}}
\def\msdlr{m_{\sdlr}}
\def\msslr{m_{\sslr}}
\def\mch{m_{H^{+}}}
\def\mA{m_{A}}
\def\mH{m_{H}}
\def\st{\widetilde{t}}
\def\sb{\widetilde{b}}
\def\mstaul{m_{\stau_{1}}}
\def\mstauh{m_{\stau_{2}}}
\def\ddf{{\rm d}}
\def\bino{\widetilde{B}}
\def\higgsino{\widetilde{H}^{0}}
\def\sz1{{\widetilde{Z}}_{1}}
\def\szs{{\widetilde{Z}}_{2}}
\def\szt{{\widetilde{Z}}_{3}}
\def\szf{{\widetilde{Z}}_{4}}
\def\szk{{\widetilde{Z}}_{k}}
\def\swl{{\widetilde{W}}_{1}}
\def\swh{{\widetilde{W}}_{2}}
\def\mse{m_{\sele}}
\def\mser{m_{\ser}}
\def\msz1{m_{\sz1}}
\def\mszs{m_{\szs}}
\def\mszt{m_{\szt}}
\def\mszf{m_{\szf}}
\def\mswl{m_{\swl}}
\def\mswh{m_{\swh}}
\def\mbino{m_{\bino}}
\def\gev{{\rm GeV}}
\def\tev{{\rm TeV}}
\def\rs{{\sqrt{s}}}
\def\tanbe{\tan\beta}
\def\nle{{\stackrel{<}{\sim}}}
\def\nge{{\stackrel{>}{\sim}}}
\def\goto{\rightarrow}
\def\bhd{{\hat{\beta'}}}
\def\half{{\frac{1}{2}}}
\def\mt{m_{t}}
\def\mb{m_{b}}
\def\mh{m_{h}}
\def\stl{\st_{1}}
\def\stls{\st^{*}_{1}}
\def\sth{\st_{2}}
\def\sbl{\sb_{1}}
\def\sbr{\sb_{R}}
\def\sbh{\sb_{2}}
\def\mst{m_{\st}}
\def\mstl{m_{\stl}}
\def\msth{m_{\sth}}
\def\msb{m_{\sb}}
\def\msbl{m_{\sbl}}
\def\msbh{m_{\sbh}}
\def\mz{m_{Z}}
\def\mw{m_{W}}
\def\tht{\theta_{t}}
\def\thb{\theta_{b}}
\def\tew{\theta_{W}}
\def\cbar{\overline{c}}
\def\sw{\sin^{2}\theta_{W}}
\def\cw{\cos^{2}\theta_{W}}
\def\muo{\mu_{\infty}}
\def\mgo{M_{\infty}}
\def\afo{A_{\infty}}
\def\mfo{m_{\infty}}
\def\mgut{M_{X}}
\def\mtil{\widetilde{m}}
\def\Xtil{\widetilde{X}}
\def\misEt{\slash\hspace{-8pt}E_{T}}
\def\pos{\slash\hspace{-6pt}p_{1}}
\def\pts{\slash\hspace{-6pt}p_{2}}
\def\prs{\slash\hspace{-6pt}p_{3}}
\def\mot{m_{12}^{2}}
\def\mor{m_{13}^{2}}
\def\mtr{m_{23}^{2}}
\def\mssql{\mtil^{2}_{Q_{1,2}}}
\def\mssqh{\mtil^{2}_{Q_{3}}}
\def\mssul{\mtil^{2}_{U_{1,2}}}
\def\mssuh{\mtil^{2}_{U_{3}}}
\def\mssdl{\mtil^{2}_{D_{1,2}}}
\def\mssdh{\mtil^{2}_{D_{3}}}
\def\mssl{\mtil^{2}_{L}}
\def\msse{\mtil^{2}_{E}}
\def\msh1{\mtil^{2}_{H_{1}}}
\def\msh2{\mtil^{2}_{H_{2}}}
\def\Ktil{\widetilde{K}}
\def\Itil{\widetilde{I}}
\preprint{ITP-SU-94/03, RUP-94-08}
\begin{title}
Light Scalar Top and Heavy Top Signature at CDF
\end{title}
\begin{author}
{Tadashi Kon}
\end{author}
\begin{instit}
Faculty of Engineering, Seikei University, Tokyo 180, Japan
\end{instit}
\begin{author}
{Toshihiko Nonaka}
\end{author}
\begin{instit}
Department of Physics, Rikkyo University, Tokyo 171, Japan
\end{instit}
\vskip5pt
\begin{center}
{(May 19, 1994)}
\end{center}
\begin{abstract}
We propose a mechanism which could explain a slight excess of
top signal rate recently reported by CDF
in the framework of the supersymmetric standard model.
If the scalar partner of the top (stop) is
sufficiently light, the gluino with an appropriate mass
could decay into the stop plus the top
with almost 100\% branching ratio and
experimental signatures of the gluino pair production could be
indistinguishable from those of the top production
in the present integrated luminosity Tevatron running.
In this case the standard top signal, $W$ $+$ multi-jets events, would
be effectively enhanced by the additional gluino contribution.
It is shown, moreover, that such a mechanism can actually work
in the radiative SU(2)$\times$U(1) breaking model
without the GUT relations between the gaugino mass parameters.
\end{abstract}
\narrowtext

The CDF experiment at the Fermilab Tevatron $p\bar{p}$ collider
has recently reported the first evidence for a top quark signal,
indicating a mass $\mt$ $=$ $174\pm 10^{+13}_{-12}$ GeV \cite{CDFtop}.
They have found an excess of $W$ $+$ multi-jets events containing at
least one $b$-jet, where the $W$ is identified by $W\to\ell\nu$ decay.
The observed signal rate
($\sigma_{t\bar{t}}$ $=$ 13.9 $^{+6.1}_{-4.8}$ pb \cite{CDFtop})
is slightly higher than those expected from the standard
$p\bar{p}\to t\bar{t}X$ cross section \cite{laenen}.
Although this higher rate may be attributed to
statistical fluctuations or background uncertainties,
some mechanisms have already been proposed
to explain this discrepancy \cite{Hill,Eichten,Barger}.
In this letter we present another mechanism
which could explain such a higher top signal rate
in the framework of the supersymmetric (SUSY)
standard model \cite{Nilles}.

In the framework of SUSY standard model with minimal
contents of ordinary particles
and their SUSY partners \cite{Nilles},
a large cross section of the gluino $\gluino$ pair production
in $p\bar{p}$ collisions is expected
because the production occurs via the QCD interaction \cite{DEQ}.
Two types of sub-processes contribute to the production, i.e.,
the gluon-gluon fusion $gg\to\gluino\gluino$ and the
quark-antiquark scattering $q\bar{q}\to\gluino\gluino$.
The latter contains the squark exchange diagrams and in turn
those contributions depend on squark masses $\msq$.
In Fig.~1 we show the gluino mass $\msg$ dependence of the total
cross section at Tevatron $\rs=1.8\tev$.
It should be noted that the cross section increases as
$\msq$ increases \cite{BTW}.
This is due to a reduction of destructive interference between
the production diagrams involving initial-state quarks.
We can obtain the total cross section $\sigma$ $\simeq$ 5~pb
for $\msg\simeq180\gev$ [$200\gev$] when $\msq=0.3\tev$ [$1\tev$].
Note that $\sigma\simeq$5pb
is almost the same value with the total cross
section of $p\bar{p}\to t\bar{t}X$ for $\mt\simeq170\gev$
expected by the standard QCD \cite{laenen}.

Searches for the gluino with large $\misEt$ signature
have been performed at CDF \cite{CDFg} and D0 \cite{D0g}
and negative results have been used to set the lower mass
bounds such as $\msg\nge 100\gev$ for $\msg\le\msq$.
In those analyses the cascade decays
$\gluino\to q\bar{q}\zino_{2,3,4}$ and
$\gluino\to ud\wino_{1,2}$ as well as the direct decay
$\gluino\to q\bar{q}\zino_{1}$
have been taken into account \cite{BTW,cascade}.

Here we consider the case $\mt+\mstl<\msg<\msq$,
where $\stl$ and $\squark$ respectively denote
the lighter scalar partner of the top (stop) \cite{stop,HK}
and all the other squarks.
Then the two-body decay mode \cite{Baer}
\begin{equation}
\gluino\to t\stls, \ \bar{t}\stl
\end{equation}
will dominate over the conventional three-body decay modes.
We assume that
({\romannumeral1})
the stop has a small mass and
the dominant decay mode $\stl\to c\sz1$ \cite{HK} and
almost degenerates in mass with the lightest neutralino
$\mstl\nge\msz1$ and
({\romannumeral2})
the top decays dominantly into $bW$ \cite{Baerf}.
In this case experimental signatures of the gluino pair production
$p\bar{p} \to \gluino\gluino X \to t\bar{t}\stl\stls X$
will resemble to the standard top signal
because the transverse momentum of the charm-jets from the final stops
will be too soft to be detected.
Moreover,
if the gluino mass is smaller than about 200GeV,
the top (like) signal rate will be effectively enhanced
since $\sigma$($p\bar{p}\to\gluino\gluino X$) $\simeq$
$\sigma$($p\bar{p}\to t\bar{t}X$) as mentioned above.
That is, the slight excess of the top signal rate at CDF
could be explained by such an additional gluino contribution.
Note that
it is required that the stop should be sufficiently light
$\mstl\nle$30GeV, owing to the conditions,
$\mstl < \msg-\mt$, $\msg \nle 200\gev$ and $\mt \sim 170\gev$.

In Fig.~2 we show various event distributions for the processes
$p\bar{p}$ $\to$ $\gluino\gluino X$
$\to$ $t\stl X$ $\to$ $bW X$ $\to$ $b\ell X$
and
$p\bar{p}$ $\to$ $t\bar{t} X$
$\to$ $bW X$ $\to$ $b\ell X$,
where we take $\mt=170\gev$, $\msg=190\gev$, $\msq=350\gev$,
$\mstl=15\gev$ and $\msz1=13\gev$.
We can find that it would be difficult for us to distigush the
gluino events from the standard top events in the present
integrated luminosity $L=19.3$pb$^{-1}$ Tevatron running \cite{CDFtop}.
It is worth mentioning that higher statistics will
enable us to confirm or reject our scenario.
Since the gluino can decay into both final states
$t\stls$ and $\bar{t}\stl$ due to its Majorana nature,
like sign $W$-boson pairs, i.e., like sign lepton pairs
will be expected in the final state of
$p\bar{p}\to\gluino\gluino X$
with an equal event rate to opposite sign lepton pairs.

Next we should examine a possibility for the existence of
a very light stop $\mstl\nle$30GeV and
a light neutralino $\msz1\nle\mstl$ \cite{ours1,ours2}.
Difficulties in the possible detection of
such a light stop through the direct searches $p\bar{p}\to\stl\stls X$ at
Tevatron \cite{Baer,Tata} as well as through the measurement of
the extra decay width of $Z$-boson
$\Gamma(Z\to\stl\stls)$ at LEP \cite{DH}
have already been pointed out.
Okada \cite{bsg} has investigated
possible bounds on masses of the stop and the neutralino
from the experimental data of the $b\goto s\gamma$ decay.
He has shown that the light stop with mass $\mstl\nle$20GeV
has not been excluded by the data.
It has been also pointed out by Fukugita et al. \cite{FMYY} that
the existence of the light stop does not conflict with the
experimental bounds on $\Delta\rho$ and $K^{0}$ $-$ $\bar{K}^{0}$
mixing.
Recently, severe limits come from the direct
searches for the stop at $e^+e^-$ colliders
\cite{VENUST,Kobayashi,DELPHI}.
In particular, OPAL group \cite{Kobayashi}
has shown the light stop $\mstl<\mz/2$ survives
only if $\mstl-\msz1<2.2\gev$ and $0.85<\tht<1.15$,
where $\tht$ denotes the mixing angle of stops \cite{HK,DH}.

It should be remarked that there is a possible sign of
the existence of a light stop.
Recently Enomoto et al. in the TOPAZ group at TRISTAN
have reported a slight excess of the high $p_T$
cross section of $D^{*\pm}$-meson production
in a two-photon process \cite{Enomoto}.
The disagreement between the measured value and the standard
model prediction becomes $3.5\sigma$ level \cite{EnomotoT}.
Although there remains a possibility that such a excess could be
explained by the large contribution of the gluonic structure
of the photon \cite{resolve},
there is another exciting way to interpret this enhancement, i.e.,
it is the pair production of the stop with mass $\nle$ $20\gev$.
Since such a light stop will decay into the charm-quark plus the
lightest neutralino \cite{HK},
the signature of the stop production will be the
charmed meson production with large missing energies.
This signature would resemble the charmed-hadron production
in the two-photon process at $e^+e^-$ colliders.
Enomoto et al. have pointed out that the stop with mass about
$15\sim16\gev$ and the neutralino with mass about
$13\sim14\gev$ could explain the
experimental data \cite{EnomotoT}.

We have examined previously \cite{ours1,ours2} such
a light stop scenario in the
framework of the minimal supergravity GUT model (MSGUT)
\cite{sugra}.
It has been emphasized that the existence of the light neutralino
$\msz1\nle20\gev$ inevitably lighten the gluino $\msg\nle100\gev$
\cite{Hidaka}
if we take the GUT relations between the soft gaugino mass
parameters at the weak scale,
\begin{eqnarray}
&&M_{1} = {\frac{5}{3}}\tan^2\tew M_{2}, \\
&&\msg = M_{3} = {\frac{\alpha_{s}}{\alpha}}\sw M_{2},
\end{eqnarray}
which are related to the boundary conditions for the
soft gaugino masses at the GUT scale,
$M_{3}(\mgut)$ $=$ $M_{2}(\mgut)$ $=$ $M_{1}(\mgut)$.
In this case the gluino can not decay into the stop plus the top
and the mechanism giving the higher event rate of the top
signal at CDF does not work.
However, if we generalize the boundary conditions for the
gaugino masses,
\begin{equation}
f_{3}^{-1}M_{3}(\mgut) = f_{2}^{-1}M_{2}(\mgut) =
f_{1}^{-1}M_{1}(\mgut),
\label{cond}
\end{equation}
the GUT relations at the weak scale are modified as \cite{ours2}
\begin{eqnarray}
&&M_{1} = {\frac{5}{3}}\tan^2\tew {\frac{f_1}{f_2}}M_{2}, \\
&&\msg=M_{3}={\frac{\alpha_{s}}{\alpha}}\sw {\frac{f_3}{f_2}}M_{2},
\end{eqnarray}
where $f_i$ are arbitrary constant parameters.
We can obtain arbitrary large gluino mass taking $f_3/f_2$ $>$ $1$.

In Fig.~3 we show
contours for $\msz1$ $=$ $13\gev$ and $\mswl$ $=$ $45\gev$
in ($\mu$, $\msg$) plane, where
we take $\tanbe$ $=$ $3$, $f_2$ $=$ $2$ and $f_3$ $=$ $2.5$,
for example \cite{f1}.
We can find that
$\msz1$ $\simeq$ $13\gev$ corresponds to
$\msg$ $\simeq$ $200\gev$ for $\mu$ $\nle$ $-50\gev$ and
such a light neutralino
has not been excluded by the limit $\mswl$ $>$ $45\gev$ at LEP
in this example.
Of course, there is a large number of parameter sets
($f_i$, $\mu$, $\tanbe$, $\msg$) which allow the existence
of the heavy gluino $\msg\nle200\gev$ and the light neutralino
$\msz1\nle20\gev$.

We can present, moreover, solutions for the renormalization group
equations (RGEs) \cite{RGE}
in the radiative SU(2)$\times$U(1) breaking model,
which satisfy all requirements mentioned above as well as
the present experimental constraints.
The RGEs and their analytical solutions
for the general boundary conditions (\ref{cond})
can be found in
Appendices A and B in Ref.\cite{ours2}, respectively.
In our calculational scheme \cite{Hikasa,ours1,ours2}
all physics at the weak scale $\mz$ are determined by the six
input parameters ($f_2$, $f_3$, $M_2$, $\mu$, $\tanbe$, $\mt$).
As an example, we take
$f_2=2$, $f_3=2.5$, $M_2=42\gev$, $\mu=-336\gev$, $\tanbe=3$ and
$\mt=170\gev$ \cite{f2}.
In this case output masses and mixing angles are
$\msg=188\gev$, $\mstl=15\gev$, $\tht=1.1$, $\msq=330\gev$,
$\msl=285\gev$, $\msz1=13\gev$, $\mszs=51\gev$, $\mswl=51\gev$,
$\mh=79\gev$ and $\alpha=-0.36$,
where $\alpha$ denotes the Higgs mixing angle \cite{GH}.
Following properties of the parameter set should be noted. \\
({\romannumeral1})
The gluino can be produced with the cross section
$\sigma$($p\bar{p}\to\gluino\gluino X$) $\simeq$
$\sigma$($p\bar{p}\to t\bar{t}X$) and can decay into
the stop plus the top, $\msg>\mstl+\mt$. \\
({\romannumeral2})
Since the stop almost degenerates with the lightest neutralino
$\mstl-\msz1<2.2\gev$,
its experimental signal, the charm-jets from $\stl\to c\sz1$
decay, would be difficult to be observed at Tevatron. \\
({\romannumeral3})
The branching ratios of the top \cite{topdc,Baer} are
BR($t\to bW$) $\simeq$ 90\% and
BR($t\to \stl\sz1, \stl\szs$) $\simeq$ 10\%. \\
({\romannumeral4})
The stop mixing angle $\tht=1.1$ lies in the range $0.85<\tht<1.15$
and the light stop $\mstl<\mz/2$ has not been excluded by
the recent direct searches at LEP \cite{Kobayashi,DELPHI}
in this case. \\
({\romannumeral5})
Masses of all the other SUSY particles
and the Higgs bosons lie in the experimentally
allowed range \cite{L3,visZ,LEPh},
where we have included the radiative correction in the calculation
of the Higgs masses \cite{higgsmass}. \\
Note that the parameter set presented here is an example and
there is a large number of parameter sets
($f_2$, $f_3$, $M_2$, $\mu$, $\tanbe$, $\mt$)
which have similar interesting properties.

In summary,
we have proposed a mechanism which could explain a slight excess of
top signal rate recently reported by CDF
in the framework of the supersymmetric standard model.
If the stop is light enough $\mstl\nle30\gev$,
the gluino with mass about $200\gev$
could decay into the stop plus the heavy top $\mt\simeq170\gev$
with almost 100\% branching ratio and
experimental signatures of the gluino pair production will be
indistinguishable from those of the top production
in the present integrated luminosity Tevatron running.
In this case the standard top signal, $W$ $+$ multi-jets events, will
be effectively enhanced by the additional gluino contribution.
It has been shown that such a mechanism could actually work
in the radiative SU(2)$\times$U(1) breaking model
without the GUT relations between the soft gaugino masses.
It shoud be emphasized that higher statistics at Tevatron
enables us to confirm or reject our scenario
through our searching for like sign lepton pairs
with an equal event rate to opposite sign lepton pairs
in the final state of the gluino pair production.

\vskip 20pt

\begin{acknowledgements}
One of the authors (T.K.) would like to thank
Y. Chikashige, I. Ito
and T. Kobayashi for helpful comments and stimulating discussions.
\end{acknowledgements}


\vfill\eject

\figure{
Gluino mass dependence of total cross section for
$p\overline{p}$ $\to$ $\gluino\gluino X$
at Tevatron $\rs=1.8\tev$.
}
\figure{
Various event distributions for
$p\overline{p}$ $\to$ $\gluino\gluino X$
$\to$ $t\stl X$ $\to$ $bW X$ $\to$ $b\ell X$
and
$p\overline{p}$ $\to$ $t\overline{t} X$
$\to$ $bW X$ $\to$ $b\ell X$.
We take $\mt=170\gev$, $\msg=190\gev$, $\msq=350\gev$,
$\mstl=15\gev$, $\msz1=13\gev$ and $\rs=1.8\tev$.
}
\figure{
Contours for $\msz1$ $=$ $13\gev$ and $\mswl$ $=$ $45\gev$
in ($\mu$, $\msg$) plane.
We take $\tanbe$ $=$ $3$, $f_2$ $=$ $2$ and $f_3$ $=$ $2.5$.
}
\vfill\eject


\begin{references}
\bibitem{Baerf}
Authors of Ref.\cite{Baer} have investigated the possibility
that the gluino decays into the top and the stop.
They have considered, however, the dominant decay mode of the top quark
would be $t\to \stl\sz1$ in their discussion.
\bibitem{f1}
Note that we can take $f_1=1$ without loss of generality.
\bibitem{f2}
$f_3>1$ is the sufficient condition for us to make the gluino
heavy. However, we must also take $f_2>1$ so as to
allow both existence of the light stop $\mstl\nle30\gev$ and
the heavy top $\mt\simeq170\gev$ \cite{ours2}.
\end{references}
\end{document}